\documentclass[reprint,amsmath,amssymb,aps,pre]{revtex4-2}
\usepackage{graphicx}  
\usepackage{dcolumn}  
\usepackage{bm}          
\usepackage{hyperref}  
\usepackage{color}
\usepackage{lipsum}

\usepackage{amsmath} 
\usepackage{dsfont}
\usepackage{cancel}
\usepackage{amsthm} 
\usepackage{amssymb}	
\usepackage{empheq}
\usepackage[absolute,overlay]{textpos}
\usepackage{xcolor}
\usepackage{tikz}
\usepackage{relsize}
\usepackage[utf8]{inputenc}

\begin{document}

\title{Nonmonotonic consensus transitions in \\ bounded-confidence dynamics on unbiased networks}
\author{Paolo Molignini}
\email{paolo.molignini@fysik.su.se}
\affiliation{Department of Physics, Stockholm University, AlbaNova University Center, 106 91 Stockholm, Sweden}
\date{\today} 

\begin{abstract}
We study the Hegselmann-Krause model of opinion dynamics on sparse, unbiased networks generated via Wilson’s algorithm, unveiling how network connectivity and confidence bounds jointly determine collective behavior. 
By systematically exploring the parameter space spanned by the confidence level $\epsilon$ and the mean degree density $\mu$, we construct comprehensive phase diagrams that classify the emergent steady states into different degrees of fragmentation and consensus.
We uncover a nonmonotonic re-entrant transition where increased connectivity can paradoxically suppress consensus, and show that full unanimity is unattainable at low connectivity due to structural isolation. 
Convergence times exhibit two distinct slowdowns: a finite-size, connectivity-dependent resonance near $\epsilon \sim 1/N$, and a critical peak associated with the established fragmentation-to-consensus transition. 
While the critical confidence threshold $\epsilon_c$ stabilizes near 0.2 for large system sizes, finite-size effects and sparse connectivity significantly alter the dynamics and phase boundaries in smaller populations. 
Our results offer new insights into the interplay between network topology and opinion dynamics, and highlight conditions under which increased connectivity may hinder, rather than promote, consensus.
\end{abstract}

\keywords{opinion dynamics, graph networks, Hegselmann-Krause model, consensus, polarization}

\maketitle

\section{Introduction}
Opinion dynamics has emerged as a prominent interdisciplinary field where concepts and techniques from statistical physics and complex systems are employed to model and understand simplified versions of social interactions~\cite{Castellano:2009, Galam:sociophysics-book, Fortunato:2005-4, Sirbu:2017}.
This growing area of research aims to describe how the individual opinions of a set of ``agents" evolve over time due to interpersonal influences and the influence of external factors like the media. 
A variety of models have been proposed and studied, such as the voter model~\cite{Holley:1975}, the Sznajd model~\cite{Sznajd:2000}, the Deffuant-Weisbuch model~\cite{Deffuant:2000}, and the Hegselmann-Krause model~\cite{Hegselmann-Krause:2002}, each capturing different aspects of social influence and interaction rules. 
These models have found remarkable applications in understanding the microscopic mechanisms of consensus formation, polarization, and fragmentation.

Among these, the Hegselmann-Krause (HK) model stands out as a seminal example of a bounded-confidence model with a nonlinear update rule. 
In the original HK model, agents communicate with others whose opinions lie within a given confidence level $\epsilon$ from their own, emulating the natural phenomenon of people preferentially interacting with like-minded individuals.
This construction leads to dynamics that are highly sensitive to initial conditions and interaction thresholds~\cite{Hegselmann-Krause:2002, Lorenz:2007-2}, depending on which the system can converge to a global consensus, fragment into several opinion clusters, or exhibit persistent polarization if subgroups remain mutually inaccessible. 
Moreover, the HK model possesses clear features of dynamical phase transitions, such as critical slowing down near consensus thresholds~\cite{Slanina:2011}.

The success of the original HK model has led to many extensions, most notably with integer opinions~\cite{Fortunato:2004}, continuous agents and density-based distributions~\cite{Lorenz:2007, MirTabatabaei:2014, Wedin:2015}, or agents belonging to multiple classes~\cite{Fu:2016} or with different characteristics~\cite{Ghaderi:2014}. Other generalizations with smoother update rules to preserve connectivity dynamically~\cite{Yang:2014}, with multi-dimensional opinion spaces~\cite{Fortunato:2005-2, Pluchino:2006, Lorenz:2006, Lanchier:2022}, under external influences~\cite{Hegselmann:2006, Chen:2019}, modeling interaction ambiguity using fuzzy logic~\cite{Zhao:2021}, or with state-dependent interaction graphs~\cite{MirTabatabaei:2012, Jia:2013, Su:2014, Jia:2015} have also been studied. 
Authors have also sought to unify the HK framework with other types of opinion dynamics models by constructing generalized update rules~\cite{Urbig:2004, Lorenz:2005, Schawe:2020}.

Comparatively speaking, the influence of graph topology on consensus formation in the HK model has received less attention.
Several works have considered how the structure or connectivity of the underlying graph affects the dynamics. 
For instance, Wedin \emph{et al.}~\cite{Wedin:2022, Hegarty:2016-2} studied circular topologies, while Fortunato~\cite{Fortunato:2005-3, Fortunato:2005} analyzed Barabási-Albert scale-free networks and conjectured that the consensus threshold depends only on whether the average degree diverges or remains finite as the number of actors diverges. 
Interestingly, Parasnis \emph{et al.}~\cite{Parasnis:2018} examined connected but incomplete graphs and showed that for any such graph, there exist initial conditions under which the dynamics never terminate - highlighting the fragility of consensus under even minimal connectivity loss. 
Similarly, Urbig \emph{et al.}~\cite{Urbig:2004} found that limiting the number of communicating agents during updates increases both the number of steady-state clusters and the convergence time. 
These findings support the intuitive idea that reduced communication lowers the likelihood of isolated agents adapting their opinions.

While the general idea of varying topology -- either by modifying agent connectivity or the update rule -- has been explored, most studies have focused on structured or heterogeneous graphs.
In particular, the study by Schawe \emph{et al.}~\cite{Schawe:2021} has focused on the role of network \emph{bridges} in facilitating consensus on regular lattices and scale-free networks. 
However, a systematic determination of phase diagrams, clustering information, and convergence properties of HK models on \emph{unbiased} graph topologies, in which edges are distributed without structural bias, has not been explicitly presented in the literature.
A truly random or uniformly sampled graph has no built-in tendencies like homophily, hierarchy, or community structure.
Therefore, it can act as a neutral baseline where to test the raw impact of graph connectedness on consensus dynamics, particularly in relation to the normalized edge density $\mu$~\cite{footnote3}.

In this work, we fill this gap by mapping out the full steady-state phase diagram in the $(\epsilon, \mu)$ plane for the HK model on sparse, unbiased graphs, across a range of system sizes. 
Our analysis shows that while large, highly connected networks reproduce the well-known fragmentation--consensus transition near $\epsilon_c \approx 0.2$, sparser and finite systems exhibit qualitatively different behavior. 
In particular, low connectivity suppresses unanimity, giving rise to multipolar states even at large $\epsilon$.
Conversely, at lower $\epsilon$, we uncover a counterintuitive regime in which increasing connectivity can disrupt consensus, reintroducing fragmentation. 
Additionally, convergence time analyses reveal both finite-size resonances and critical slowing down near the consensus transition.
Taken together, our findings show that bounded confidence dynamics on unbiased networks can give rise to rich and unexpected collective phenomena -- well beyond the predictions of fully connected models.

The rest of this paper is structured as follows.
In Sec.~\ref{sec:model}, we introduce the mathematical formalism of the HK model and its simply-connected graph generalization.
In Sec.~\ref{sec:quantities}, we discuss how to extract relevant information from the dynamics.
We present the main results from our numerics and discuss their implications in Sec.~\ref{sec:results}.
Finally, we offer some conclusions and general perspectives in Sec.~\ref{sec:discussion}.


\section{Model}
\label{sec:model}
We consider a generalization of the HK model~\cite{Hegselmann-Krause:2002} describing the opinion dynamics of $N$ interacting agents $\mathcal{A}_i$, $i=1, \dots, N$.
The agents' opinions $\{o_i \}$ are assumed to be univariate and constrained within the interval $[0,1]$~\cite{footnote1}.
We are interested in a procedure that makes the opinions $o_i(t)$ change over time.
For simplicity, time is treated discretely and labeled by an integer $t=0, 1, 2, \dots$.
The dynamics of each agent's opinion -- i.e. the value at the next time step -- is determined by an \emph{update rule} that depends on the opinion of its neighbors in the current time step.
More precisely, in the HK model the update rule takes the form
\begin{equation}
o_i (t+1) = \frac{1}{|\mathcal{I}_i|} \sum_{\mathcal{A}_j \in \mathcal{I}_i} o_j(t),
\label{eq:update-rule}
\end{equation}
where $\mathcal{I}_i \equiv \left\{ \left\{ \mathcal{A}_j \right\}_j \bigg| \: \mathcal{A}_j \in \mathcal{N}_i \land |o_j(t) - o_i(t)| < \epsilon \right\}$ is the set of all (nearest) neighbors $\mathcal{N}_i$ of agent $\mathcal{A}_i$ whose opinions fall within a threshold $\epsilon$ of its own.
This threshold is often denoted as \emph{confidence level} in the literature, and we will maintain this terminology here.
In this work, the confidence level is set globally, although heterogeneous versions of the model have been studied~\cite{Bernardo:2020, Perrier:2024}.
As the initial state, we will employ a uniform distribution of opinions between 0 and 1 for all cases.

Whereas in the original HK model the connections among the agents were all-to-all, in this work we will consider the agents in a simply connected graph network topology with $N$ nodes (representing the agents) and $M$ edges (representing their connections).
An example of this topology is given in Fig.~\ref{fig:model}(a).
The all-to-all connectivity of a fully connected graph topology might be useful to model small committees in which every member has a direct relationship and discusses with (and hence influences) everyone else.
However, as the number of agents $N$ progressively increases, it becomes less realistic that all agents can interact and influence each other.

Research suggests that a typical person can retain between 100 and 250 connections~\cite{Dunbar:1992}, with a typical average of 150 known as Dunbar's number.
If we sample sizes that are greater than this number, we are bound to encounter situations in which not every agent can influence everyone else.
Other additional external factors could further restrict the communication between individuals who would otherwise share the same opinion, such as language barriers, socio-economic background, geographical distance, etc.
Furthermore, even if we consider smaller sizes, very often we are interested in studying sectors of society with non-fully connected topologies, which are instead better described by entangled networks of people sharing the same environment.
Examples include schools, companies, large events, etc.
It thus makes sense to reduce the connectivity of the graph and study simply connected but not complete topologies~\cite{footnote2}.


\section{Numerics and quantities of interest}
\label{sec:quantities}
The main feature we are interested in calculating is the behavior of the system in the long-time limit, i.e. the values $o_i(t \to \infty)$.
If the opinions saturate to a final configuration with
\begin{equation}
|o_i(\tilde{t}) - o_i(\tilde{t}+n)| < \eta \quad \forall i \in \{1, 2, \dots, N\}, n \in \mathbb{N},
\label{eq:ss-criterion}
\end{equation}
with $\eta$ a small enough threshold, we call the opinion distribution at times $t>\bar{t}$ the \emph{steady state}.
In this work, we set $\eta=10^{-6}$, which for the system sizes considered ends up being a stricter threshold than previously used~\cite{Schawe:2021}.
Previous work has shown that the HK model always reaches a converged state~\cite{Blondel:2010,Bhattacharyya:2013, Wedin:2015-2}.

In practice, we are interested in three main quantities to classify the steady state: (1) the number of separate opinion clusters $C$ it exhibits, (2) their sizes $S_j$ $j \in [1, 2, \dots, C]$ (for simplicity we order the sizes from largest to smallest, i.e. $S_1>S_2> \cdots > S_C$), and (3) the convergence time $\tilde{t}$ it takes to reach the steady state.
If $C=1$, all the agents have converged to the same opinion and we refer to the corresponding steady state as a \emph{unanimous} state~\cite{Schawe:2021}.
All the other cases with $C>1$ correspond to a split in the opinion distribution.
However, in this case the size of the clusters also matters.
Two quantify this size more easily, we introduce the relative cluster size $\mathcal{R} = \frac{S_1}{\sum_{j=1}^C S_j}$.
The unanimity limit exhibits $\mathcal{R} \to 1$, whereas a complete opinion fragmentation with single-agent clusters will tend to $\mathcal{R} \to 1/N$.
If one cluster dominates in size over the others, i.e. $S_1 \gg S_j$ $\forall j>1$, we call the steady state a \emph{consensus} state.
In agreement with previous work~\cite{Schawe:2021}, we will set the threshold for consensus at 90\%. 
Moreover, we will refine the agreement pool by also introducing the notion of a (simple) \emph{majority}, in which 60 to 90\% of the agents share the same opinion ($0.6 \le \mathcal{R} < 0.9$). 
If instead multiple clusters have a comparable population, the steady state is called \emph{fragmented} or \emph{polarized}.
In our numerics, we will call a state polarized when only two clusters appear and the dominant cluster accounts for 50-60\% of the total agents ($0.5 \le \mathcal{R} < 0.6$). 
To understand the degree of fragmentation beyond 2-cluster polarization, we will also monitor the emergence of cluster states with three clusters of comparable population, i.e. where the majority cluster contains 33-50\% of all agents ($0.33 \le \mathcal{R} < 0.5$).

\begin{figure}[t!]
\centering
\includegraphics[width=\columnwidth]{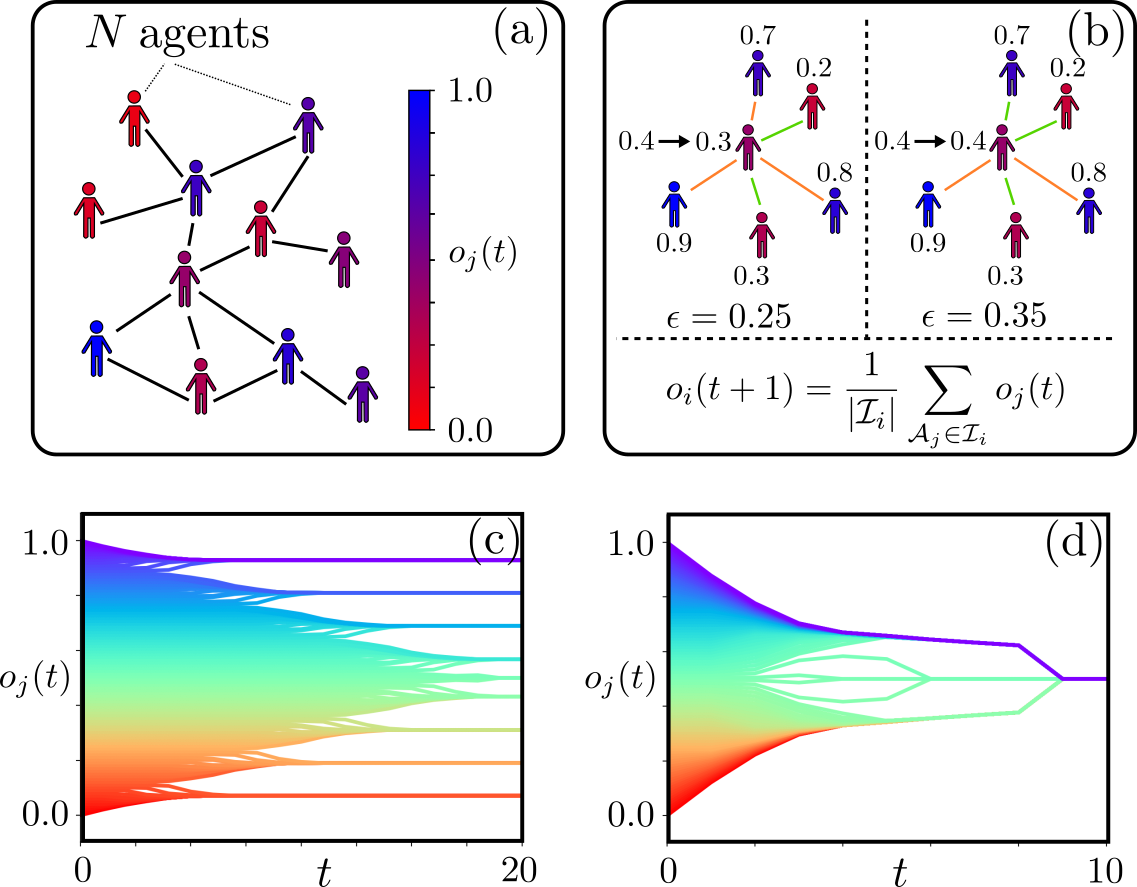}
\caption{
(a) Sketch of the generalized Hegselmann-Krause model used in this work: $N$ agents with time-dependent opinions $o_j(t) \in [0, 1]$ are located at the vertices of a simply-connected but not fully connected graph, and interact with one another.
(b) The opinions evolve according to the update rule \eqref{eq:update-rule}, which depends on both the agent's connections and the confidence level $\epsilon$. 
Only neighbors with an opinion within $\epsilon$ enter the update rule.
(c) Time evolution of a uniformly distributed state ($N=100$) below the critical confidence level $\epsilon_c$, showcasing steady-state polarization.
(d) Time evolution of a uniformly distributed initial state ($N=100$) above the critical confidence level $\epsilon_c$, showcasing steady-state consensus.
}
\label{fig:model}
\end{figure}

As we are modelling opinion dynamics in a societal context, it is thus interesting to determine which conditions lead to the emergence (or absence) of unanimity or consensus.
This can be expressed in terms of the system parameters like $N$ and $\epsilon$, but also its graph structure.
The appearance of consensus as a function of the confidence level has been amply studied~\cite{Lorenz:2010, Hegarty:2016}.
Generally speaking, at small values of $\epsilon$, the opinions in the steady state converge to a multitude of separate clusters, whereas for larger values of $\epsilon$ the number of steady state clusters is reduced until unanimity is reached, as shown in Fig.~\ref{fig:model}(c)-(d), respectively.

The threshold for consensus $\epsilon_c$ depends on the number of agents.
For fully connected topologies, $\epsilon_c(N)$ decreases slowly with increasing system size, saturating at $\epsilon_c = 0.2$ in the large $N$ limit depending on initial conditions and averaging conventions~\cite{Fortunato:2005}, although finite-size effects can shift the apparent critical value.
For instance, in the fully-connected case with $N=100$ consensus appears around $\epsilon_c=0.25$. 
In complex or sparse networks, significantly larger values of $\epsilon$ up to $\epsilon_c = 0.5$ may be required to ensure unanimity~\cite{Fortunato:2005}.
Consensus, instead, can be reached with arbitrarily small $\epsilon$ in large scale-free networks, owing to the high probability of finding bridges linking clusters of different opinions~\cite{Schawe:2021}.

In our numerics, we simulate the time evolution according to the update rule \eqref{eq:update-rule} from an initial uniformly-populated opinion space.
We consider systems with $N=50$, $100$, $200$, and $500$ agents.
To study how the transition between consensus and polarization is modified by the graph's topology, we simulate $N_r=1000$ random realizations of simply connected graphs with normalized edge density $\mu$, defined as $\mu = \frac{M - M_{\mathrm{min}}}{M_{\mathrm{max}} - M_{\mathrm{min}}} = \frac{2(M-N+1)}{N(N-3) + 2}$.

To guarantee that each graph is unbiased, we construct each graph from its uniform spanning tree via Wilson's algorithm~\cite{Wilson:1996}, and then we populate it with additional edges until the desired value of $\mu$ is reached.
A summary of Wilson's algorithm is provided in appendix~\ref{app:wilson}.
We monitor the opinion convergence for each time evolution, and once a steady state according to the criterion in Eq.~\eqref{eq:ss-criterion} is reached, we count the number of clusters and their population in the final opinion distribution.
Given the large number of simulations due to the randomization of the graph's topology, we automate the cluster counting with an unsupervised learning algorithm.
For the results shown in this work, we have employed a KMeans algorithm~\cite{MacQueen:1967}, but we have verified that other algorithms like kernel density estimation~\cite{Rosenblatt:1956} lead to similar results.
We probe steady-state properties such as mean cluster number, convergence time, and relative cluster size as a function of two main parameters: the confidence level $\epsilon$, and the mean edge density $\mu$.

\section{Results}
\label{sec:results}

\begin{figure}[t!]
\centering
\includegraphics[width=\columnwidth]{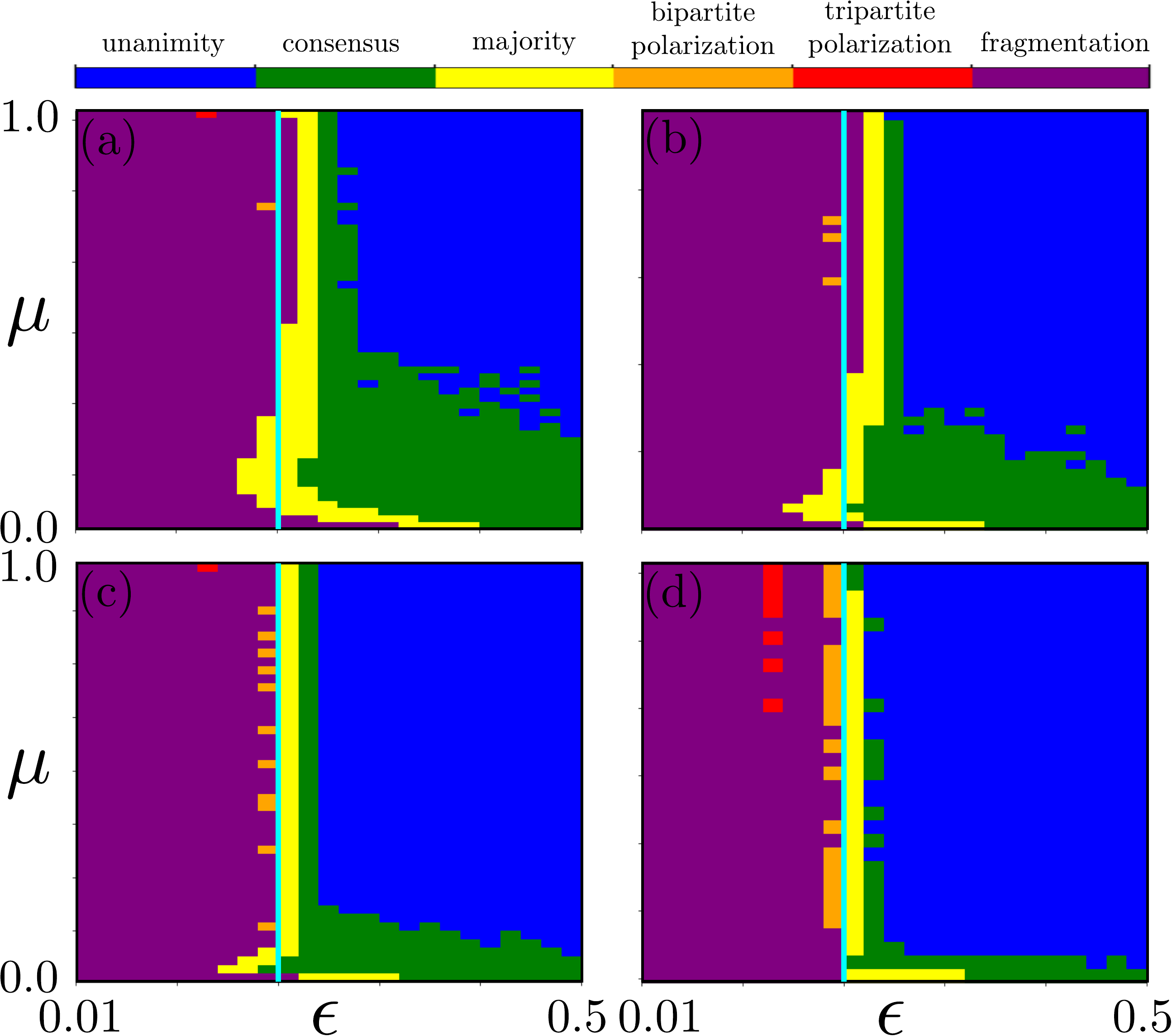}
\caption{
The different steady-state phases appearing in a Hegselmann-Krause model on unbiased graphs, in the parameter space spanned by the confidence level $\epsilon$ and the mean edge density $\mu$.
Different panels present results for different system sizes: 
(a) $N=50$,
(b) $N=100$,
(c) $N=200$,
(d) $N=500$.
}
\label{fig:phase-diagram}
\end{figure}

\begin{figure}[t!]
\centering
\includegraphics[width=\columnwidth]{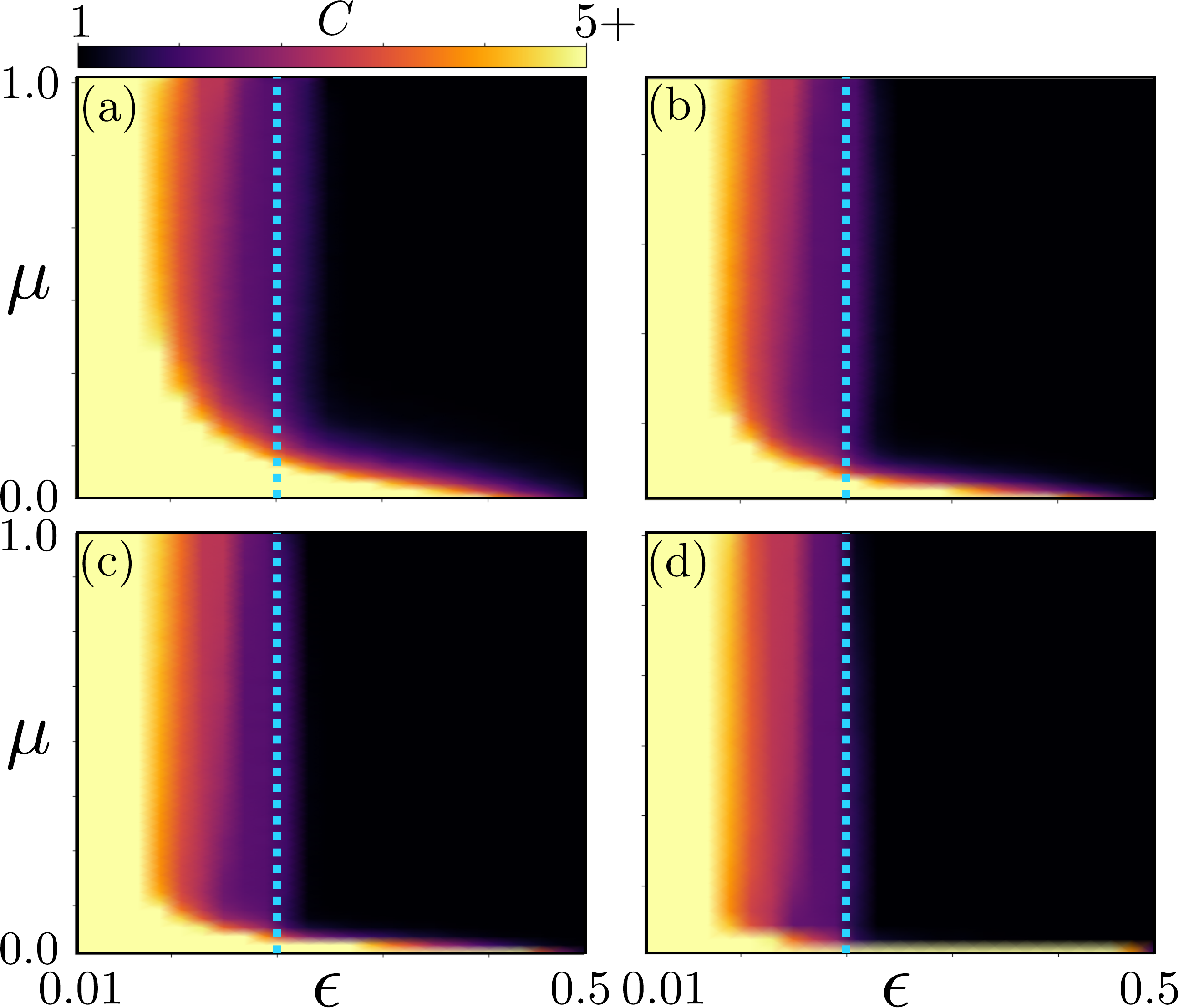}
\caption{
The number of steady-state clusters in the parameter space spanned by the confidence level $\epsilon$ and the mean edge density $\mu$.
Different panels present results for different system sizes: 
(a) $N=50$,
(b) $N=100$,
(c) $N=200$,
(d) $N=500$.
}
\label{fig:clusters}
\end{figure}

\begin{figure}[t!]
\centering
\includegraphics[width=\columnwidth]{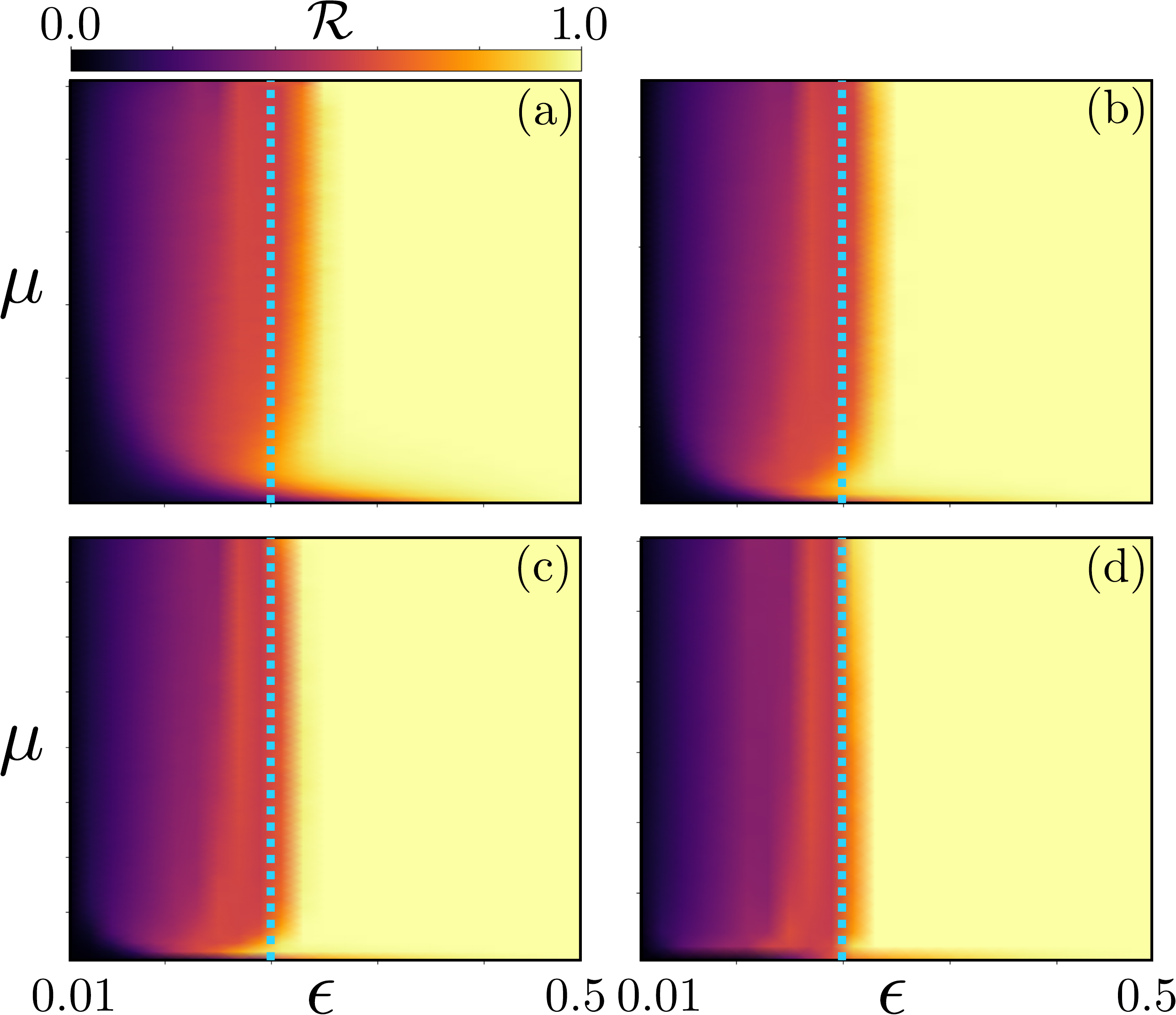}
\caption{
The relative size $\mathcal{R}$ of the largest steady-state cluster in the parameter space spanned by the confidence level $\epsilon$ and the mean edge density $\mu$.
Different panels present results for different system sizes: 
(a) $N=50$,
(b) $N=100$,
(c) $N=200$,
(d) $N=500$.
}
\label{fig:cluster-size}
\end{figure}

We now present the results of our simulations.
Fig.~\ref{fig:phase-diagram} displays the full phase diagram in the $(\epsilon,\mu)$ plane for all system sizes probed, revealing the rich structure of steady-state opinion configurations in the HK model on unbiased graphs
Each region of the diagram is color-coded according to the emergent collective state: unanimity (blue), consensus (green), majority (yellow), bipartite polarization (orange), tripartite polarization (red), and generic fragmentation (purple), as defined in Sec.\ref{sec:quantities}.
This phase diagram is compiled using the number of steady-state clusters shown in Fig.\ref{fig:clusters} and the relative size of the largest cluster presented in Fig.~\ref{fig:cluster-size}.

As expected, fragmentation dominates at low confidence levels, while unanimity emerges at high $\epsilon$.
However, the transition between these two regimes is highly sensitive to the average connectivity $\mu$ of the network.

For large $\mu$, the transition is remarkably sharp.
For instance, at $N=50$, the system evolves from unanimity to consensus, then to majority, and finally to fragmentation, over a narrow window $\Delta \epsilon \approx 0.07$ centered near $\epsilon \approx 0.25$.
As $N$ increases, the transition shifts toward the conjectured thermodynamic limit $\epsilon_c \approx 0.2$~\cite{Fortunato:2005, Schawe:2020, Schawe:2021} and becomes increasingly abrupt, with thinner regions of consensus or majority.

For low $\mu$, the behavior is radically different.
The system no longer reaches unanimity, even at high $\epsilon$.
Instead, the dominant state is a consensus state, in which one dominant opinion encompasses over 90\% of agents, but smaller clusters can coexist.
Moreover, the width of the transition region grows dramatically, reaching $\Delta \epsilon \approx 0.33$.
This consensus region is dynamically stable; however, it shrinks as $N$ increases, suggesting that larger systems can achieve unanimity by partially compensating for sparse connectivity because the initial opinion distribution is denser.
Interestingly, bipartite and tripartite polarization are rare and typically only occur at larger $N$.

A particularly intriguing feature appears around the critical value $\epsilon_c=0.2$ for low connectivity ($\mu < 0.15$), where a clear re-entrant behavior of the consensus and majority phases is observed. 
This region can be analyzed from two perspectives.
At fixed $\epsilon$, increasing $\mu$ initially facilitates consensus, but further increases lead paradoxically to fragmentation.
This nonmonotonic behavior is rather counterintuitive, since in a more interconnected network it should be easier to find neighbors with similar opinions to converge to the same final opinion.
This re-fragmentation effect instead suggests that excessive connectivity can reinforce local opinion islands and hinder global consensus -- a fundamentally non-monotonic response absent from classical HK models.
On the other hand, at fixed low connectivity $\mu$, consensus can emerge at much lower values than the thermodynamic limit threshold $\epsilon_c=0.2$ when operating at finite system sizes.
For example, with $N=200$ agents and $\mu=0.05$, a majority can be achieved with as little as $\epsilon=0.13$.
These features demonstrate the subtle and unintuitive phenomenology of opinion dynamics in complex, partially connected topologies.

\begin{figure}[t!]
\centering
\includegraphics[width=\columnwidth]{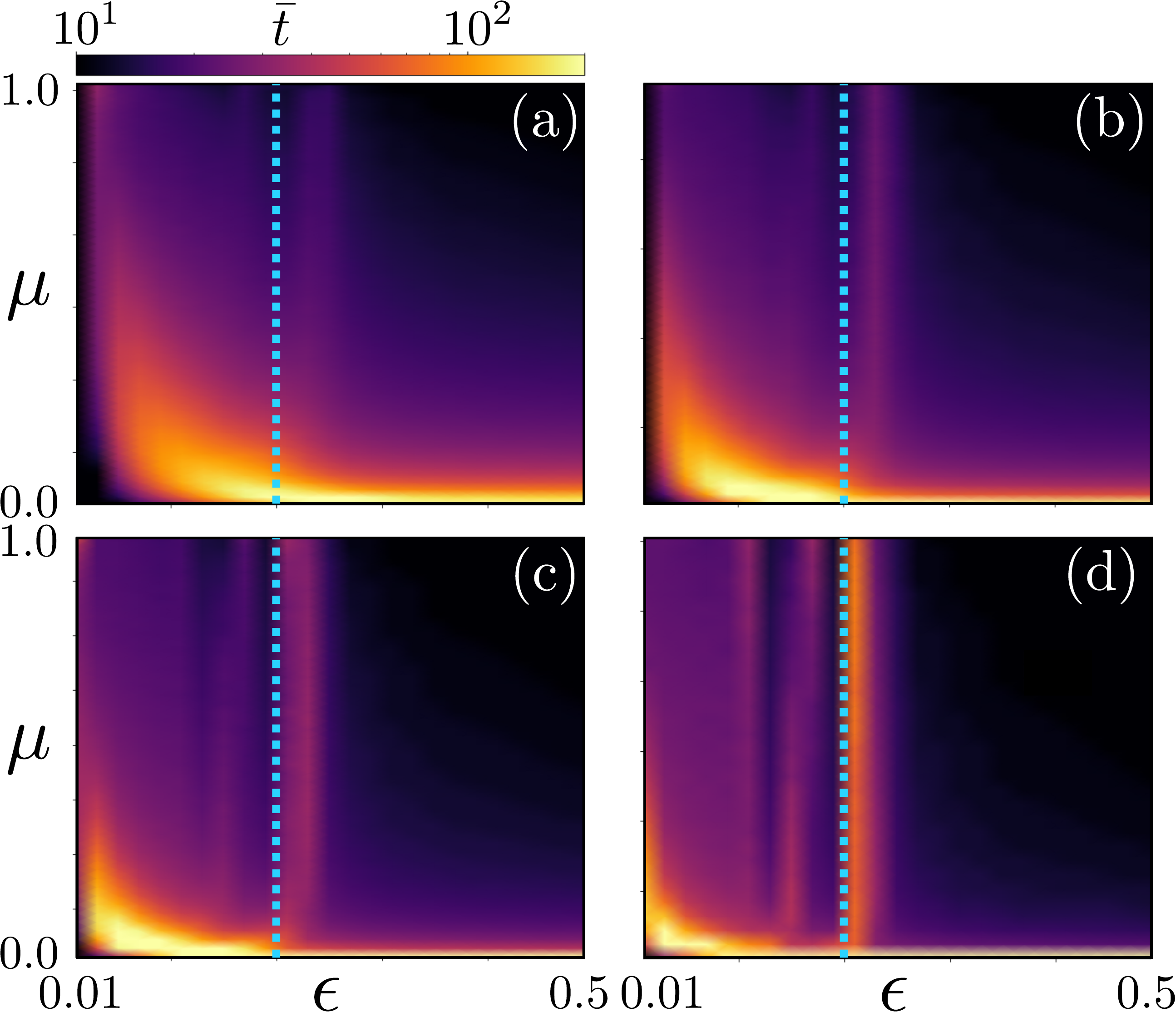}
\caption{
The average time needed to reach the steady state as a function of confidence level $\epsilon$ and the mean edge density $\mu$.
Different panels present results for different system sizes: 
(a) $N=50$,
(b) $N=100$,
(c) $N=200$,
(d) $N=500$.
}
\label{fig:conv-time}
\end{figure}

\begin{figure}[t!]
\centering
\includegraphics[width=\columnwidth]{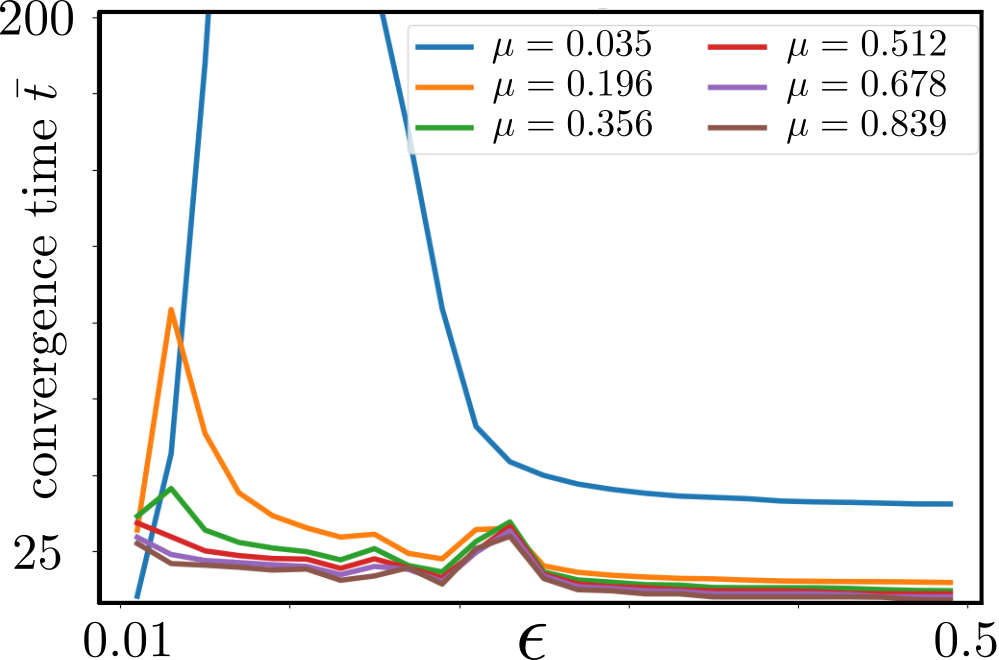}
\caption{
The average time needed to reach the steady state as a function of confidence level $\epsilon$ for a few selected values of $\mu$ for $N=200$ agents.
}
\label{fig:conv-time-cuts}
\end{figure}

Additional interesting observations can be made from the average convergence times for our simulations.
These are presented in Fig.~\ref{fig:conv-time}
At fixed edge density, convergence time exhibits two clear peaks: a first, stronger one in the fragmented region $\epsilon < \epsilon_c$, and a second, weaker one in concomitance with the transition to consensus at $\epsilon_c$.
These features are similar to the results obtained in scale-free networks~\cite{Schawe:2021}.
This picture is reiterated in the cuts at fixed $\mu$ shown in Fig.~\ref{fig:conv-time-cuts}.
This double peak structure results in a nonmonotonic convergence time as a function of confidence level $\epsilon$.
Away from these peaks, fragmentation (below $\epsilon_c$) or consensus (above $\epsilon_c$) instead are reached rather quickly.

The height and location of the first peak strongly depends on the edge density $\mu$, with higher peaks occurring for smaller values of $\mu$ and larger values of $\epsilon$. 
This results in a hyperbolic shape traced in the $(\epsilon,\mu)$ plane [cf. Fig.~\ref{fig:conv-time}].
This feature is a finite-size effect, linked to a resonant configuration where agents are just at the edge of mutual influence: for a given number of agents $N$ there exists a value of the confidence threshold $\epsilon ~ \frac{1}{N}$ for which the initial distance among equally spaced opinions is such that each agent (except the two at the opinion boundaries) will have exactly two neighbors a distance $\epsilon$ away, essentially cancelling out the effect of the update rule and leaving all bulk agents' opinions almost unaffected.
In this regime, agents effectively become inert, slowing down convergence.
This is especially pronounced at low $\mu$, where sparse connections exacerbate isolation and prevent opinion updates from propagating efficiently.
As $N$ increases, the critical $\epsilon \sim 1/N$ shifts toward zero, squeezing the peak closer to the origin.

The second peak, on the other hand, remains robust as $N$ increases and reveals the location of the phase transition from fragmentation to consensus.
While the transition appears smeared at finite $N$ -- due to the effect of decreasing the connectivity -- it sharpens as $N$ grows, as larger populations effectively average out local deviations and restore a mean-field-like behavior.
This is also reflected in the convergence time associated with this peak, consistent with the critical slowing down expected near a second-order transition~\cite{Slanina:2011}.

Altogether, our findings indicate that while low connectivity can strongly affect finite systems by enabling re-entrant behavior and smearing unanimity, the core features of the phase transition remain robust in the thermodynamic limit.
Importantly, the critical threshold $\epsilon_c \approx 0.2$ emerges universally across topologies, including the important unbiased cases studied in this work.

\section{Discussion and outlook}
\label{sec:discussion}

Our study uncovers a complex and nuanced landscape for opinion formation in the Hegselmann-Krause model on sparse, unbiased networks.
By systematically exploring the full $(\epsilon,\mu)$ parameter space, we demonstrate that the emergence of collective agreement is jointly governed by both confidence level and network connectivity.
This results in important deviations from the predicted behavior of the fully-connected Hegselmann-Krause dynamics in low-connectivity environments, including smearing of consensus, re-entrant phases, and non-monotonic behavior.
System size plays here a crucial modulating role, pushing these feature to progressively lower connectivity as the number of interacting agents increases.

One of the central findings is the sharpening and convergence of the critical threshold $\epsilon_c$ toward 0.2 as the system size increases, in agreement with classical HK results for fully connected graphs.
However, in sparse networks, this transition becomes broadened and shifted, especially at small $\mu$, where full unanimity is essentially inaccessible due to structural isolation.
Instead, consensus and majority states emerge as more robust forms of partial agreement, where a dominant opinion coexists with minor dissenting clusters.

The loss of unanimity at low $\mu$ is accompanied by an intriguing re-entrant fragmentation at intermediate connectivities, indicating that overly connected structures can paradoxically inhibit global coordination.
We interpret this as a topological frustration effect: sparse links encourage local convergence but prevent global reconciliation when initial opinions are evenly distributed. 

The phase diagrams also reveal that bipartite and tripartite polarization are rare and likely require both large $N$ and specific conditions -- suggesting that these highly structured divisions may be emergent features of larger-scale societies or due to other interplays between agents.

Convergence time mapped across parameter space exposes two distinct slowdowns: a finite-size-induced plateau associated with resonant configurations near $\epsilon \sim 1/N$, and a robust critical peak near $\epsilon_c$ signaling the fragmentation-consensus transition.
The former becomes negligible in the thermodynamic limit, as observed in other kinds of graphs -- while the latter sharpens with increasing $N$, echoing the hallmarks of second-order phase transitions.

Our results highlight the intricate and often unpredictable dependencies between individual bounded confidence, global topology, and collective behavior.
They suggest that interventions aimed at increasing connectivity in real-world networks may have nontrivial outcomes, potentially aiding or hindering consensus depending on the regime and on the scale at which they operate.

Several promising avenues emerge from this work.
First, it would be valuable to characterize the network structure of emergent opinion clusters -- e.g., whether they correspond to modules in the interaction graph, or whether they are spatially delocalized.
Second, introducing stochasticity (e.g., opinion noise or asynchronous updates) could test the stability of the observed phases.
Similarly, analyzing opinion resilience under perturbations (e.g., targeted rewiring or removal of influential agents) may yield insight into the robustness of the wide consensus state that dominates at low connectivity.
Third, a comprehensive study of the role of the initial states should be performed to reveal how the different pathways to consensus and fragmentation can be impacted.
For example, while dense links appear to enhance unanimity for larger confidence levels when starting from a uniform opinion distribution, they may preserve local divisions and reinforce fragmentation if initial opinions are sufficiently polarized.
Finally, it would be instructive to consider directed or time-evolving graphs, especially in light of applications to social media or organizational dynamics.

\acknowledgments 
We thank Supriya Krishnamurthy for useful discussions.
This work was supported by the Swedish Research Council (2018-00313) and Knut and Alice Wallenberg Foundation (KAW) via the project Dynamic Quantum Matter (2019.0068).

\appendix
\section{Wilson's algorithm}
\label{app:wilson}
In order to study the Hegselmann-Krause model on randomly generated, simply connected graphs, it is crucial to start from a graph-generation method that ensures a uniform distribution over all graphs with a given number of nodes and edges. 
In this section, we present a brief overview of the algorithm used to generate such graphs.

A common source of bias arises when a spanning tree~\cite{Pemantle:1991} -- the ``skeletons'' of the graph -- is constructed with heuristics that bias towards certain topologies (e.g., star graphs) over others.
The algorithm first presented by Wilson~\cite{Wilson:1996} solves this problem by generating uniformly random spanning trees using loop-erased random walks~\cite{Lawler:1980, Lyons-Peres}.

The process begins by generating a uniform spanning tree with $N$ nodes and $N-1$ edges.
Wilson’s algorithm works as follows:
\begin{enumerate}
\item Choose an arbitrary root node.
\item For each remaining node \emph{not} in the tree, perform a loop-erased random walk:
 perform a random walk over all nodes until it reaches any node already in the tree. 
During the walk, keep track of the path and erase loops as they form.
Once the path connects to the existing tree, add all its loop-erased edges to the tree.
\end{enumerate}
This process continues until all nodes are connected. 
The resulting structure is a uniformly sampled spanning tree, free from degree or path biases present in naive constructions.

Once a uniform spanning tree is obtained, we can reach the desired total number of edges $M$ with $M > N - 1$, by incrementally adding random edges to the tree. 
At each step, we select a pair of distinct nodes uniformly at random and add an edge if it is not already present in the graph.
Self-edges are also discarded.
This approach preserves connectedness and gradually increases the graph’s density. 

Since the base tree is sampled uniformly and additional edges are added uniformly at random, the final graph remains an unbiased sample from the ensemble of connected graphs with $N$ nodes and $E$ edges. 
This ensures that the structural properties of the network are not skewed toward any particular structure, an essential condition for robust exploration of opinion dynamics in the Hegselmann-Krause model.

Wilson’s algorithm operates in expected time proportional to the cover time of the graph. 
For dense graphs (like complete graphs), the cover time is typically $\mathcal{O}(N \log N)$~\cite{Wilson:1996}, making the method practical for the moderate-sized systems studied in this work.
Edge addition is linear in the number of added edges, assuming constant-time edge existence checks via adjacency sets.


\begin{thebibliography}{99}
\bibitem{footnote3} Note that here we are calculating the average connections as $M/N$ without doubling the number of edges.
\bibitem{footnote1} The size of the interval is not a particularly important variable here, and it is selected between 0 and 1 to facilitate analytical and mostly numerical calculations. For instance, the interval $[-1,1]$ could also be employed, but cancellations between negative and positive values might arise during normalization, which complicates the numerics.
\bibitem{footnote2} Disconnected graphs have different groups of agents that never interact with agents outside of their own groups, and are therefore not particularly interesting to model opinion dynamics.
\end{thebibliography}
\end{document}